# Evaluation of the hydrogen solubility and diffusivity in proton-conducting oxides by converting the PSL values of a tritium imaging plate

M. Khalid Hossain [1,2,*], K. Hashizume [1], Y. Hatano [3]

[1]*Department of Advanced Energy Engineering Science, Interdisciplinary Graduate School of Engineering Science, Kyushu University, Kasuga, Fukuoka 816-8580, Japan.*

[2]*Atomic Energy Research Establishment, Bangladesh Atomic Energy Commission, Savar, Dhaka 1349, Bangladesh.*

[3]*Hydrogen Isotope Research Center, Organization for Promotion of Research, University of Toyama, Toyama 930-8555, Japan.*

Correspondence: *khalid@kyudai.jp; khalid.baec@gmail.com

---

## Graphical Abstract

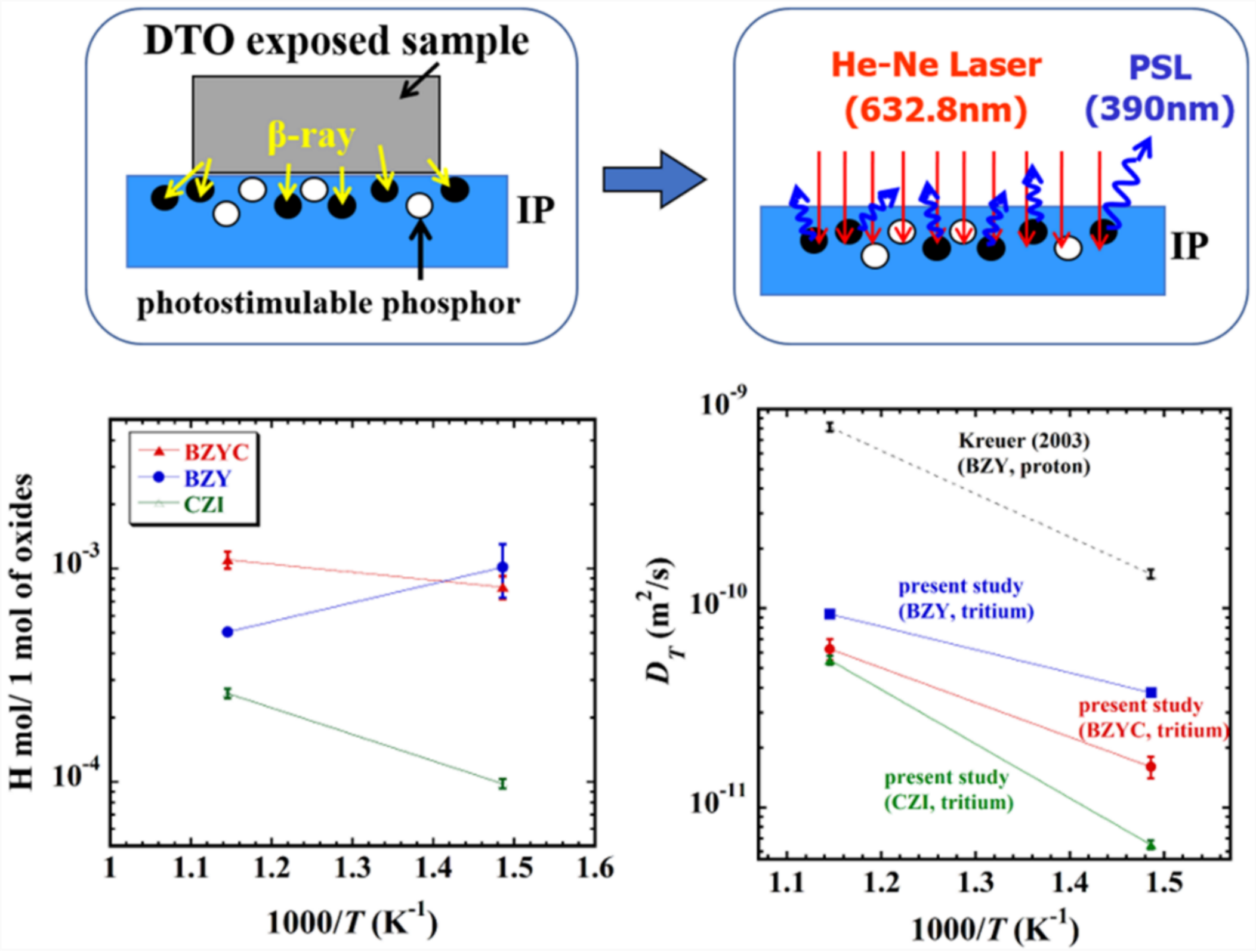

## Highlights

- Hydrogen (H) distribution in BZY, BZYC, and CZI was visualized using DTO and TIP
- DTO promoted the uniform dissolution of tritium throughout the BZY sample
- At 673–873 K, tritium diffused deeper into BZY and BZYC than into CZI
- The H solubility and diffusivity in CZI was lower than that of BZY and BZYC
- The activation energy of H-diffusion for CZI was approximately double of that for BZY and BZYC

# 1 Abstract

Proton-conducting oxides have potential applications in hydrogen sensors, hydrogen pumps, and other electrochemical devices including the tritium purification and recovery systems of nuclear fusion reactors. Although the distribution of hydrogen (H) in such oxide materials is an important aspect, its precise measurement is difficult. In the present study, the hydrogen solubility and diffusivity behavior of $BaZr_{0.9}Y_{0.1}O_{2.95}$ (BZY), $BaZr_{0.955}Y_{0.03}Co_{0.015}O_{2.97}$ (BZYC), and $CaZr_{0.9}In_{0.1}O_{2.95}$ (CZI) were studied using tritiated heavy water vapor i.e., DTO (~2 kPa, tritium (T) = 0.1%) by converting the photostimulated luminescence (PSL) values of the imaging plate (IP). The samples were exposed to DTO vapor at 673 K for 2 h or at 873 K for 1 h. The disc-shaped oxide specimens (diameter ~7.5 mm; thickness ~2.3 mm; theoretical density (TD) > 98 %) were prepared by conventional powder metallurgy. The IP images of the specimen surfaces of all the three materials T-exposed revealed that BZY showed the most uniform T distribution with the highest tritium activity. The cross-sectional T concentration profiles of the cut specimens showed that T diffused deeper into BZY and BZYC than into CZI. The hydrogen solubility and diffusivity in the CZI specimen were lower than that in the BZY and BZYC specimens. This suggested that barium zirconates were more favorable proton conductors than calcium zirconates.

**Keywords:** Barium zirconate ($BaZrO_3$); calcium zirconate ($CaZrO_3$); photostimulated luminescence (PSL); tritium imaging plate, hydrogen solubility and diffusivity, fusion reactors materials.

# 2 Introduction

A closed deuterium-tritium (DT) fuel cycle is used in the international thermonuclear experimental reactor (ITER). Here, a minor portion of D and T is used as fuel while a major portion is ejected as He-purge gas. T and D along with the He-purge gas passes through a hydrogen pump or an isotope separation system where it is purified to produce D and T that are reinjected as a fuel into the ITER fuel loop [1]. Proton-conducting oxides are potential materials for application in isotope separation systems [2–7] where various unpurified gases like DTO, He, $CD_4$, and $CT_4$ are allowed to pass through the proton-conducting ceramics. When DC power is applied between the two electrodes in this system, purified DT, $D_2$, and $T_2$ are generated (**Fig. 1**) that may be re-used as fuel in the ITER fuel system [5–7]. Therefore, proton-conducting oxides are vital for tritium purification and recovery in the hydrogen sensors and hydrogen pumps of fusion reactors [3,4,7,8].

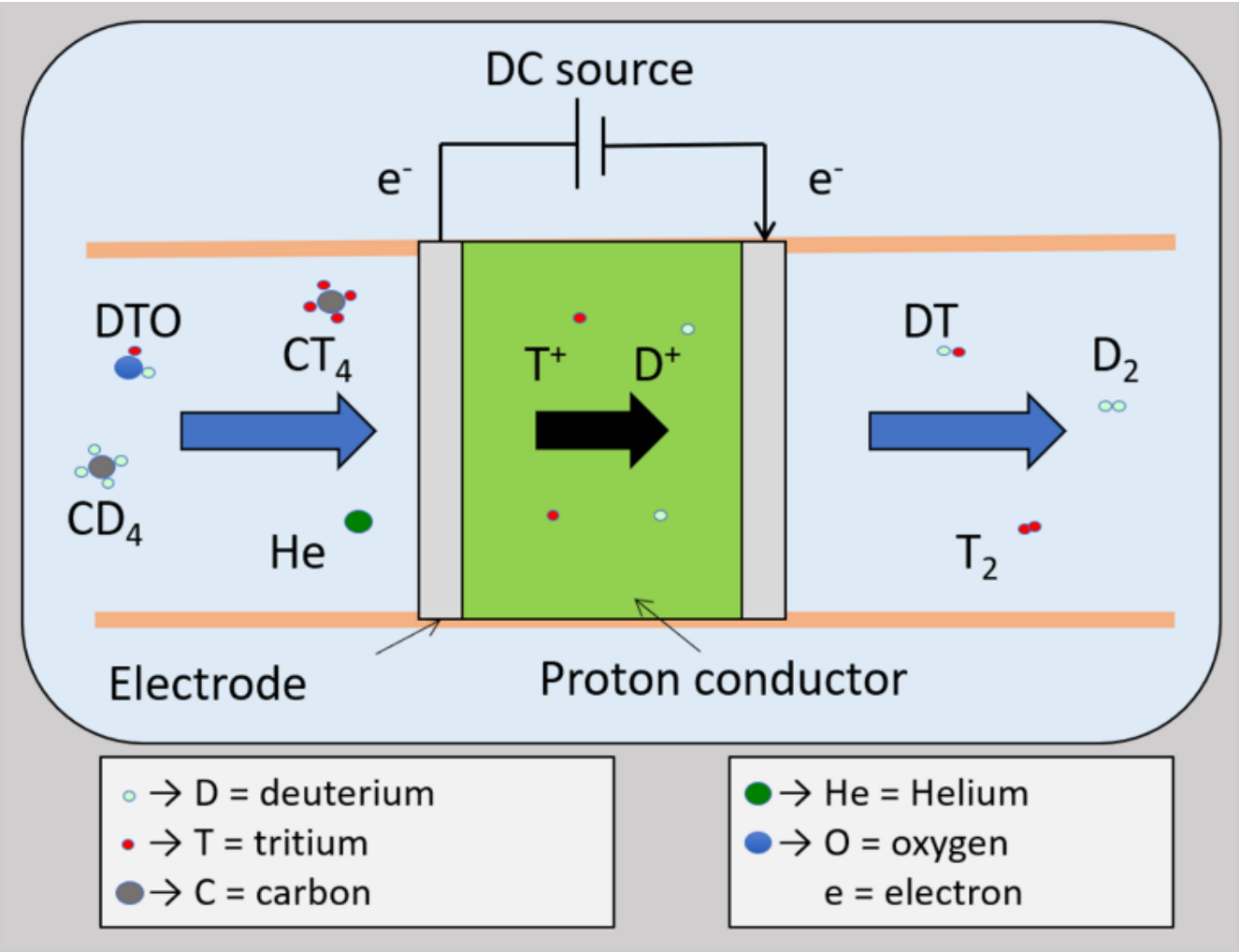

**Fig 1.** Schematic of the D-T purification system where proton-conducting oxide materials are used in isotope separators.

It is important to understand the solubility and internal diffusivity of hydrogen in proton-conducting oxides to promote the practical utilization of these materials as functional materials in fusion reactors [9–12]. Despite our previous studies on the hydrogen solubility and diffusivity behaviors of seven rare-earth oxide materials [9], the limited and scattered data for other oxide materials makes it difficult to precisely measure the hydrogen behavior. Tritium tracer methods like the tritium imaging plate (TIP) technique are powerful tools to not only measure the hydrogen concentration but also clarify the hydrogen distribution behavior in oxide materials [10–13]. The objective of this study was to obtain the hydrogen solubility and diffusivity data by converting the photostimulated luminescence (PSL) intensity of the imaging plate (IP). The IP images were used to visualize the distribution of the hydrogen isotopes in the proton-conducting oxides to determine the possible applications of these materials in fusion reactors.

Perovskite oxides are lattice defect-type proton-conducting oxides that have the potential for applications in fusion reactors due to their high-temperature operational capabilities [14–16]. Barium zirconate ($BaZrO_3$) and calcium zirconate ($CaZrO_3$), among multiple perovskite oxides, are potential materials for applications in fusion reactors due to their high chemical stabilities, proton conductivities, hydrogen pumping abilities, tritium recovery efficiencies, and hydrogen sensing capabilities [17–28]. The proton conductivities of $BaZrO_3$ and $CaZrO_3$ can be increased by partially substituting the Zr component with trivalent cations ($Y^{3+}$, $In^{3+}$, $Yb^{3+}$, and $Gd^{3+}$). $Y^{3+}$-doped $BaZrO_3$ ($BaZr_{1-x}Y_xO_{3-\alpha}$) exhibits the highest proton conductivity [29] that can be further increased by increasing the doping amount of Y up to $x$ = 0.2 i.e., 20% [30,31]. In $BaZrO_3$, the replacement of two $Zr^{4+}$ ions with two $Y^{3+}$ ions generates an oxygen vacancy (**Fig. 2(a)**). These oxygen vacancies increase the proton conductivities of the yttrium-doped barium zirconates. Some research groups have investigated the possible applications of 10% Y-doped [18,32,33], 15% Y-doped [34], and 20% Y-doped [35] $BaZrO_3$ as electrolyte materials for hydrogen sensors and hydrogen pumps. The poor sinterability of yttrium-doped barium zirconates necessitates the addition of sintering aids to enhance the grain growth [36,37]. The sinterability and catalytic activity in $BaZr_{1-x}Y_xO_{3-\alpha}$ can be increased by partially substituting the Zr components with 3d transition metals like Fe, Co, and Ni [38,39]. M. Tanaka and T. Ohshima [21] proposed the co-doping of $BaZrO_3$ with Y and Co to form $BaZr_{0.955}Y_{0.03}Co_{0.015}O_{3-\alpha}$ that has the potential for applications in fusion reactors due to its excellent hydrogen pumping capabilities at intermediate temperatures (700 K). $CaZrO_3$ exhibits high proton conductivity and excellent electrochemical properties when the Zr component is partially substituted with 10% $In^{3+}$ [40–42]. $CaZrO_3$ doped with In possesses high chemical stability and sinterability [43,44]; furthermore, its mechanical strength, density, and thermal shock resistance is higher than that of other protonic conductors. Therefore, In-doped $CaZrO_3$ is a potential candidate material for hydrogen sensors, tritium monitors, tritium pumps, and tritium recovery in fusion reactors [19,20,22–28,45–47].

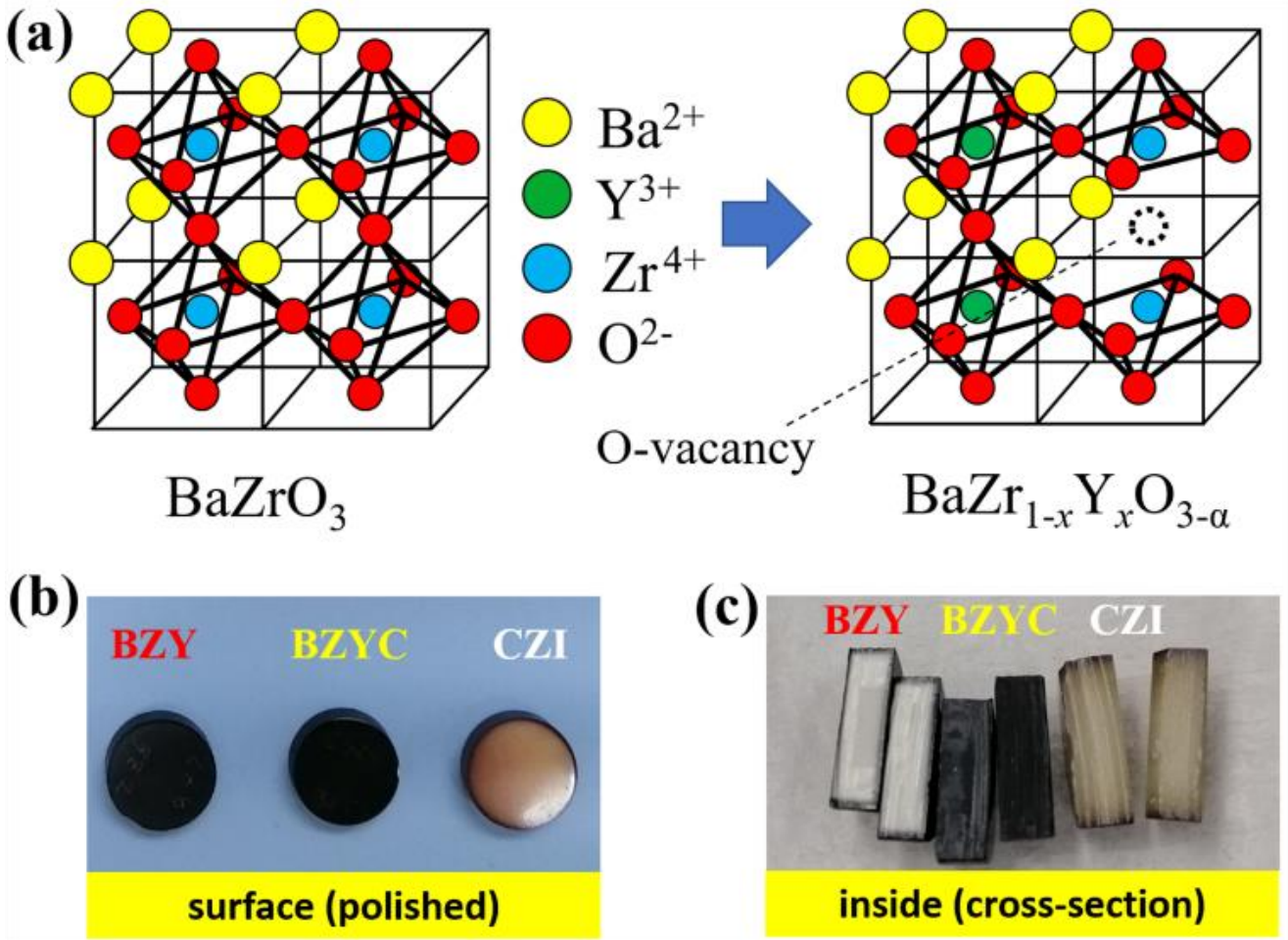

**Fig 2.** (a) Schematic of the generation of an oxygen vacancy by the substitution of two $Zr^{4+}$ ions with two $Y^{3+}$ ions in perovskite $BaZrO_3$, a proton-conducting oxide, to form $BaZr_{1-x}Y_xO_{3-\alpha}$. (b) Surface view of the sintered BZY, BZYC, and CZI samples after mirror polishing and (c) cross-sectional (inside) view of the sintered BZY, BZYC, and CZI samples after cutting.

Therefore, we selected the Y- and Co-doped barium zirconates ($BaZr_{0.9}Y_{0.1}O_{2.95}$ and $BaZr_{0.955}Y_{0.03}Co_{0.015}O_{2.97}$) and the In-doped calcium zirconate ($CaZr_{0.9}In_{0.1}O_{2.95}$) as the specimens for this study. A detailed comparison of the hydrogen (tritium) behavior in the barium and calcium zirconates was presented in this study. We described the procedure to calculate the hydrogen solubilities from the PSL values of the IP through comparisons with the ART-123A tritium standard. We also described the procedure to calculate the hydrogen diffusivities from the tritium concentration profile of the cross-sectional IP images.

# 3 Experimental

## 3.1 Sample preparation

Calcined powders of $BaZr_{0.9}Y_{0.1}O_{2.95}$ (BZY), $BaZr_{0.955}Y_{0.03}Co_{0.015}O_{2.97}$ (BZYC), and $CaZr_{0.9}In_{0.1}O_{2.95}$ (CZI) (TYK Corporation, Japan) were used as the raw materials. The sintered pellets of BZY, BZYC, and CZI were prepared by a solid-state reaction method [9,13,48–50]. 0.7 g of these powders was uniaxially compressed (load = 0.5 ton) in molds (diameter = 10 mm) to form disk-shaped samples. These disk-shaped samples were placed in a rubber tube and subjected to cold isostatic pressing (200 MPa). The samples were loaded into an alumina tube furnace with a molybdenum (Mo) heater and sintered at 1913 K for 20 h in an air atmosphere. The diameter and thickness of the sintered samples were approximately 7.5 mm and 2.7 mm, respectively. The apparent densities of the sintered samples were more than 98% of the theoretical density. X-ray diffraction (XRD) analysis confirmed the single phase of the cubic structure for all the sintered samples. The surfaces of the sintered samples were wet-polished with waterproof SiC abrasive papers and finished by mirror polishing. The diameter, thickness, and weight of the polished samples were approximately 7.5 mm, 2.0–2.3 mm, and 0.6 g, respectively. The surface and cross-sectional views of the polished samples are shown in **Figs. 2(b)** and **(c),** respectively.

## 3.2 Tritium exposure

The BZY, BZYC, and CZI samples were exposed to T-D water vapor or DTO vapor to facilitate the absorption of tritium into the samples. **Fig. 3** shows a schematic view of the DTO-exposure apparatus at the Hydrogen Isotope Research Center (HIRC) of the University of Toyama. The experiment was performed using DTO vapor with a tritium concentration of 0.1 %. To initiate DTO exposure, the polished samples were placed in a quartz tube, and the inside of the apparatus was evacuated to approximately $10^{-7}$–$10^{-8}$ Torr using a rotary pump and a turbo molecular pump. The impurities on the surfaces of the samples were removed by vacuum annealing at 1273 K for 1 h before DTO-exposure. Thereafter, the samples were kept at a predetermined temperature (673 K or 873 K). The sealed vacuum apparatus was filled with DTO vapor that was injected at 10 Torr. Tritium was dissolved in the sample after a predetermined exposure time (1 h or 2 h). The quartz tube containing the samples was quenched with room-temperature water after the DTO-exposure was completed to allow the dissolution of tritium in the samples. The DTO vapor that was remaining in the line was recovered by cooling the outside of the DTO tube with liquid $N_2$. Subsequently, the sample was removed. The distribution of tritium was observed by an imaging plate technique [10].

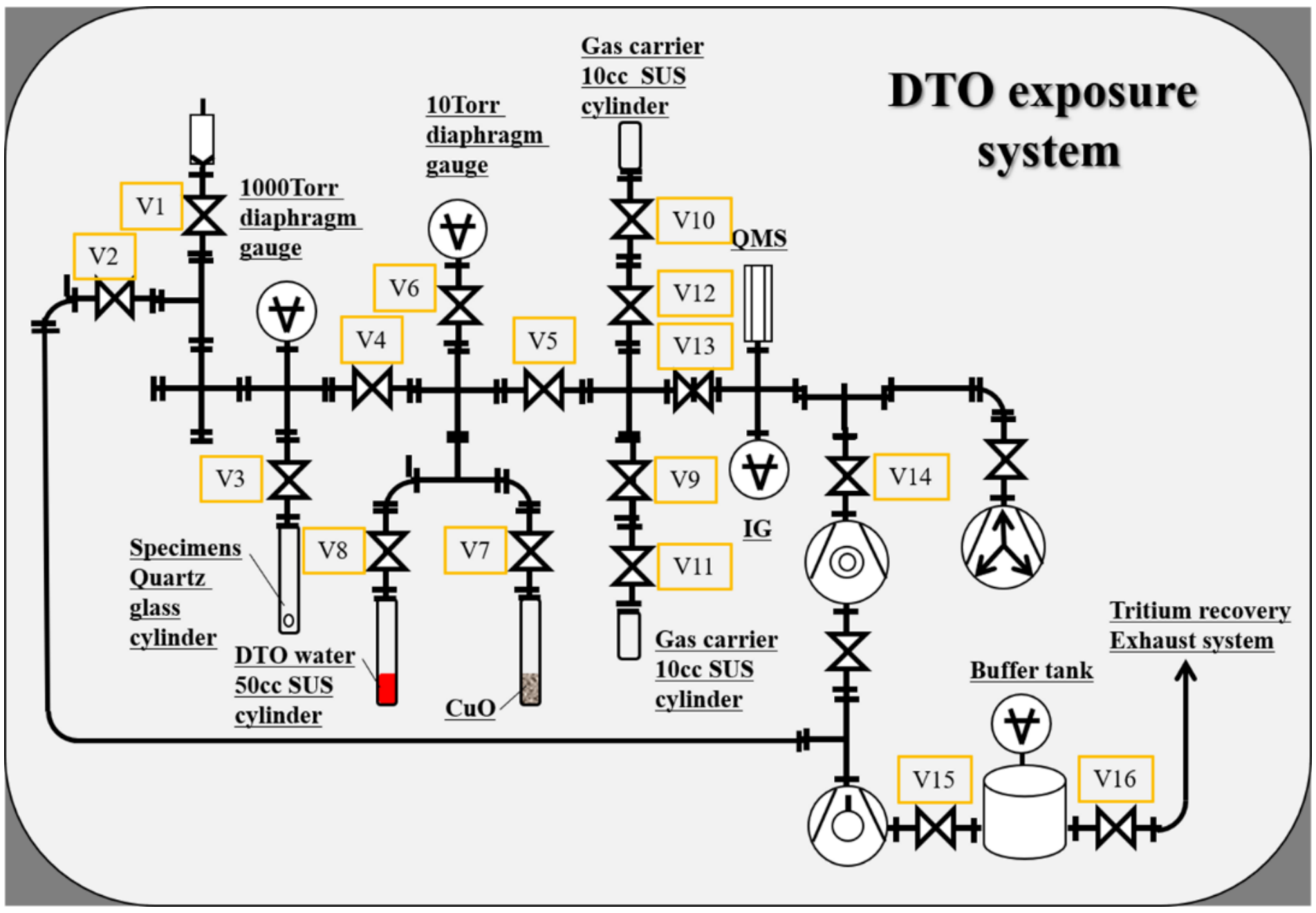

**Fig 3.** Schematic of the partially tritiated deuterium vapor i.e., DTO (T = 0.1%, ~2 kPa) exposure apparatus at the Hydrogen Isotope Research Center, University of Toyama, Japan. Here, the vacuum lines were evacuated (~$10^{-5}$ Pa) by a rotary pump and turbomolecular pump before the sample was subjected to DTO-exposure.

## 3.3 Tritium imaging plate technique

The imaging plate technique can be used to observe and quantitatively measure the radiation distribution using the PSL phenomenon of IPs. IPs are highly sensitive two-dimensional integral radiation detectors that possess an excellent linear responsiveness to radiation doses and a high position resolution. In this study, TR2010 (Fujifilm Inc., Japan) and FLA 7000 (Fujifilm Inc., Japan) were used as an imaging plate and IP reader, respectively. Since the tritium is considered to be the radiation source, the technique is called the tritium imaging plate (TIP) technique. Tritium emits a low-energy beta ray with a maximum and average energy of 18.6 keV and 5.7 keV, respectively. Since the range of the beta rays in condensed materials is low, its radiation energy is adsorbed by the photostimulable phosphor materials of the IP. Eu-doped barium fluorohalide ($BaFX:Eu^{2+}$ (X= Cl, Br, I)) is the major materials of an IP where the PSL phenomenon is used to measure the tritium. **Fig. 4** shows a schematic of the TIP technique to illustrate the principle of PSL. When the $BaFX:Eu^{2+}$ (X= Cl, Br, I) crystals adsorb the radiation energy of beta rays, few electron-hole pairs are produced. The stability of the electrons and holes contributes to the PSL phenomenon. The tritium-exposed IP was illuminated by a He-Ne laser light. This stimulating light was used to read the information on the emission position and emission intensity. Thus, the emission was detected as PSL.

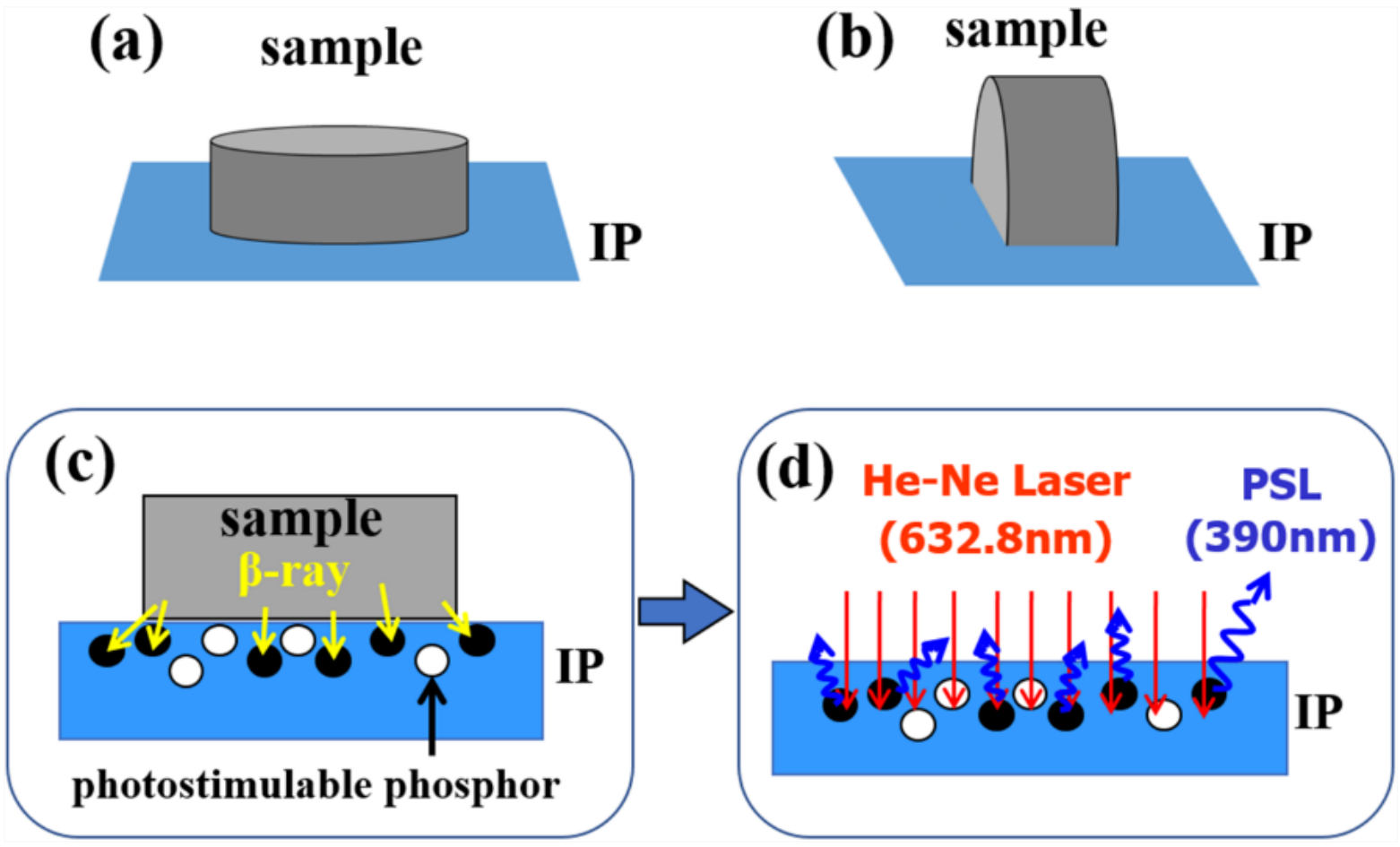

**Fig 4.** A schematic representation of the tritium imaging plate (TIP) technique showing the (a) IP exposure for the surface of the DTO-exposed sample, (b) IP exposure for the cross-section of the cut sample, (c) β-ray irradiation from tritium that is dissolved in the DTO-exposed sample to the IP, and (d) the mechanism of reading the IP information by laser irradiation.

## 3.4 Measurement of the hydrogen distribution at the surface and cross-section by IP

To capture the IP images, the surfaces of the tritium-exposed samples were placed on the IP (**Fig. 4(a)**) and stored inside a dark cassette for 1 h. The surfaces of the samples were polished and cut into two halves afterwhich their IP images were obtained. The cross-sections of the cut samples were placed on the IP (**Fig. 4(b)**) and exposed to the beta rays for 17 h. The exposure of the IP to the beta rays emitted by the tritium in the specimen is called lithographic exposure (**Fig. 4(c)**). The IP reader with red He-Ne laser light (λ= 650 nm) was used to analyze the quantity and distribution of the beta rays. The phosphor released its accumulated radiation energy by the PSL phenomenon in the form of blue light (λ= 390 nm) (**Fig. 4(d)**). The two-dimensional distribution and the incident dose irradiated from the source were determined as the emission position (the spatial resolution of FLA 7000 was 25 μm) and the emission intensity (PSL), respectively. The IP reader provided color images of the emission position and emission intensity. The PSL value was denoted as an arbitrary unit of the radiation dose; furthermore, it was proportional to the product of the time of lithographic exposure and the radioactivity of the specimen on the TIP. It was concluded that a high fluorescence intensity of PSL represented a high concentration of tritium.

# 4 Calculations

## 4.1 Conversion of the PSL value to the hydrogen solubility

The tritium activity in the sample was expressed as the PSL intensity (PSL/mm$^2$·sec). The PSL value of the standard sample varied with the IP measurement conditions, environmental conditions, and the IP sheet. Therefore, the conversion coefficient was calculated from the PSL value of the standard sample that was placed on the same IP sheet during the IP exposure. The procedure to calculate this conversion coefficient is described in this section.

As aforementioned, the IP was irradiated with beta rays that were emitted by the tritium in both the ART-123A standard and the oxide samples. Therefore, it was necessary to know the range of beta rays that were irradiated from the near-surface of the oxide sample and the tritium standard sample to calculate the amount of hydrogen. The range of beta rays of various energies in metals and ceramics was estimated by using Eqs. 1–3 that were formulated by Gledhill [51]:

$$x = \log E\, ; \quad where\ 0.1 < E < 100\ \text{keV} \tag{1}$$

$$y = \log R_p \tag{2}$$

$$y = -5.1 + 1.358x + 0.215x^2 - 0.043x^3 \tag{3}$$

Here, $E$ is the energy of the rays (in keV) and $R_p$ is the practical range (in g/cm$^2$). $E$ was considered to be 18.6 keV that is the maximum energy of the rays emitted by tritium. The densities of the prepared BZY, BZYC, and CZI samples were 6.1 g/cm$^3$, 6.16 g/cm$^3$, and 4.68 g/cm$^3$, respectively. The range of the beta rays in BZY, BZYC, and CZI was estimated to be 1.3 μm, 1.2 μm, and 1.6 μm, respectively, using the Eqs. 1–3.

The values in PSL/mm$^2$ were converted to PSL/cm$^3$·sec by considering the [$^3$H] standard beta ray range as 8.3 μm (**Table 1).** The ART-123A [$^3$H] standard sample was made of a polymer material of unknown density. We assumed the bulk density of the tritium standard sample to be 0.92 g/cm$^3$; subsequently, the radioactivity values (**Fig. 5)** in μCi/g were converted to Bq/cm$^3$ (**Table 2**). The average radioactivity of the [$^3$H] standard was estimated to be (7.71 ± 0.08) ×10$^4$ Bq· sec/PSL (**Tables 1 and 2**). This average value was used as a proportional coefficient to convert the PSL value of the samples into radioactivity to derive the amount of tritium and the hydrogen solubility.

**Table 1.** Conversion of the PSL/mm$^2$ values of the seven levels of the ART-123 [$^3$H] standard to PSL/cm$^3$·sec.

| **ART-123A [$^3$H]** | **PSL/mm$^2$** | **PSL/cm$^3$·sec** |
|---|---|---|
| Level-1 | 1738 | 3.38 x 10$^3$ |
| Level-2 | 871 | 1.69 x 10$^3$ |
| Level-3 | 466 | 9.07 x 10$^2$ |
| Level-4 | 228 | 4.44 x 10$^2$ |
| Level-5 | 111 | 2.16 x 10$^2$ |
| Level-6 | 54 | 1.05 x 10$^2$ |
| Level-7 | 30 | 5.84 x 10$^1$ |

**Table 2.** Conversion of the μCi/g values of the seven levels of ART-123 [$^3$H] to Bq/cm$^3$.

| **ART-123A [$^3$H]** | **μCi/g** | **Bq/cm$^3$** |
|---|---|---|
| Level-1 | 7390 | 2.52 x 10$^8$ |
| Level-2 | 3830 | 1.30 x 10$^8$ |
| Level-3 | 2000 | 6.81 x 10$^7$ |
| Level-4 | 1000 | 3.40 x 10$^7$ |
| Level-5 | 489 | 1.66 x 10$^7$ |
| Level-6 | 243 | 8.27 x 10$^6$ |
| Level-7 | 138 | 4.70 x 10$^6$ |

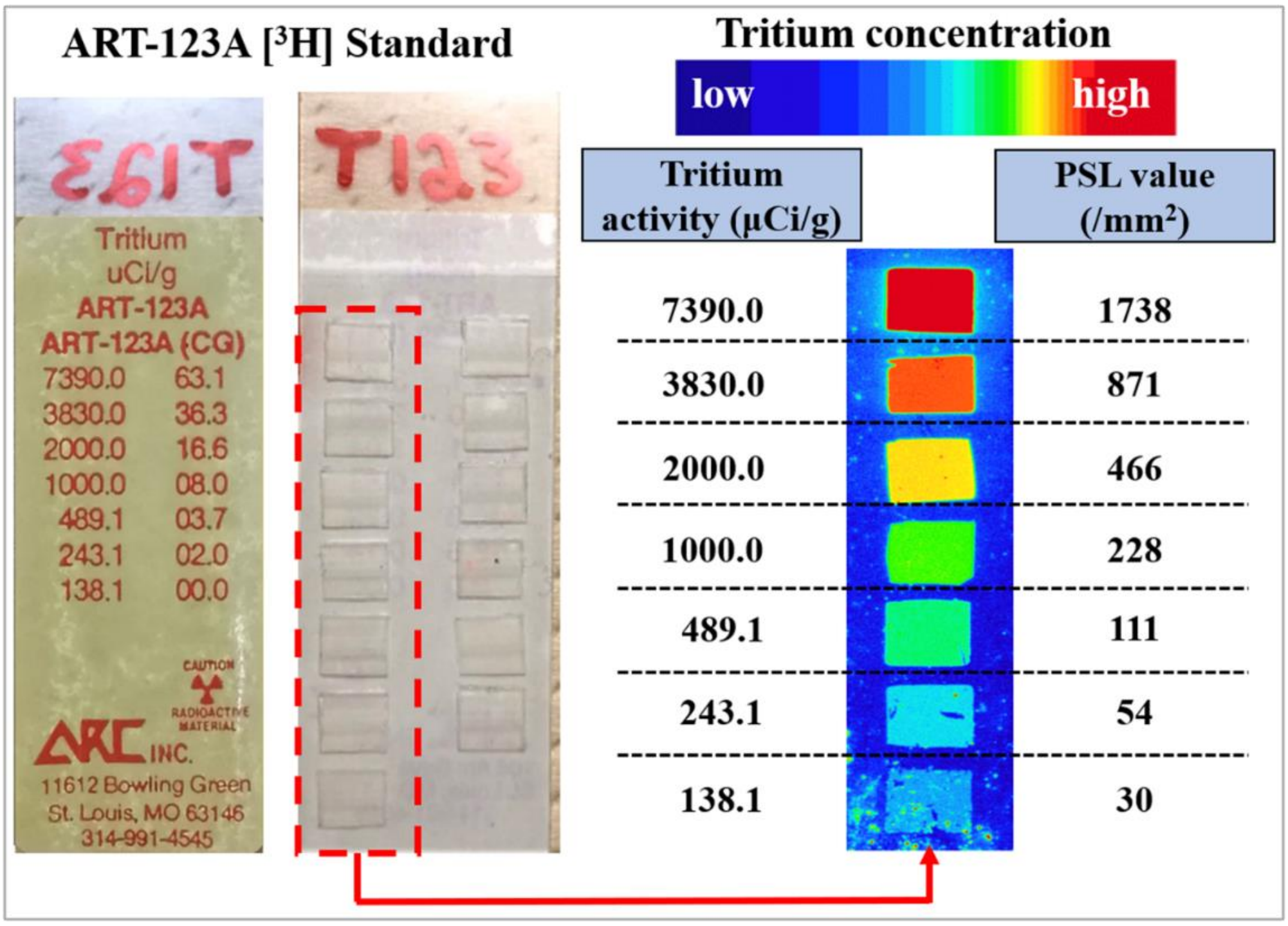

**Fig 5.** Polymer material-based ART-123A [$^3$H] standard sample on a glass substrate and its radioactivity values along with the corresponding IP images.

The BZYC sample was exposed to tritium at 673 K for 2 h. The IP measurements of this sample were performed simultaneously with that of the standard sample. The average PSL value near the surface of the BZYC sample was 3.10×10$^2$ PSL/cm$^3$·sec. This value was multiplied with the constant of proportionality to obtain the radioactivity in Bq.

$$3.10 \times 10^2\ \mathrm{PSL/cm^3 \cdot sec} \times\ 7.71 \times 10^4\ \mathrm{Bq \cdot sec/PSL} = 2.39 \times 10^7 \mathrm{Bq/cm^3}$$

The radioactivity was converted to the number of tritium moles using the radioactivity half-life formula given in Eq. 4.

$$A_{Bq} = N_A \frac{\ln\ (2)}{t_{1/2}} \quad (4)$$

Here, $A_{Bq}$ is the radioactivity in Bq, $N_A$ is the number of tritium moles (T), $t_{1/2}$ is the half-life of tritium i.e., 12.32 years. The number of tritium moles per Bq was calculated to be 5.61×10$^8$ T/Bq. Therefore,

$$2.39 \times 10^7\ \mathrm{Bq/cm^3} \times 5.61 \times 10^8\ \mathrm{T/Bq} = 1.34 \times 10^{16}\ \mathrm{T/cm^3}$$

Furthermore, the T/D ratio in the tritiated heavy water (DTO vapor) that was used in this experiment was considered to be ~10$^{-3}$.

$$1.34 \times 10^{16}\ \mathrm{T/cm^3} \times 10^3 = 1.34 \times 10^{19}\ \mathrm{D/cm^3} \cong\ 1.34 \times 10^{19}\ \mathrm{H/cm^3}$$

The amount of oxide moles per unit volume of BZY (density = 6.1 g/cm$^3$; molecular weight = 275.72 g/mol) was calculated to be 2.21 × 10$^{-2}$ mol/cm$^3$. The Avogadro number (6.02 × 10$^{23}$) was used and the number of oxide moles (M) per unit volume of BZY was calculated to be 1.33 × 10$^{22}$ M/cm$^3$. Thus, the total number of hydrogen moles per unit volume was calculated to be $1.34 \times 10^{19}\ \mathrm{H/cm^3}$ ≅ 1.01 × 10$^{-3}$ H/M. This value represented the solubility of hydrogen in the BZY oxides at 673 K after 2 h of DTO-exposure. Therefore, the number of dissolved hydrogen moles in per mole of the BZY oxides was 1.01 × 10$^{-3}$. We calculated the hydrogen solubilities for all the samples from the PSL/mm$^2$ values of the IP images (**Figs. 6** and **7)** by following the abovementioned procedure.

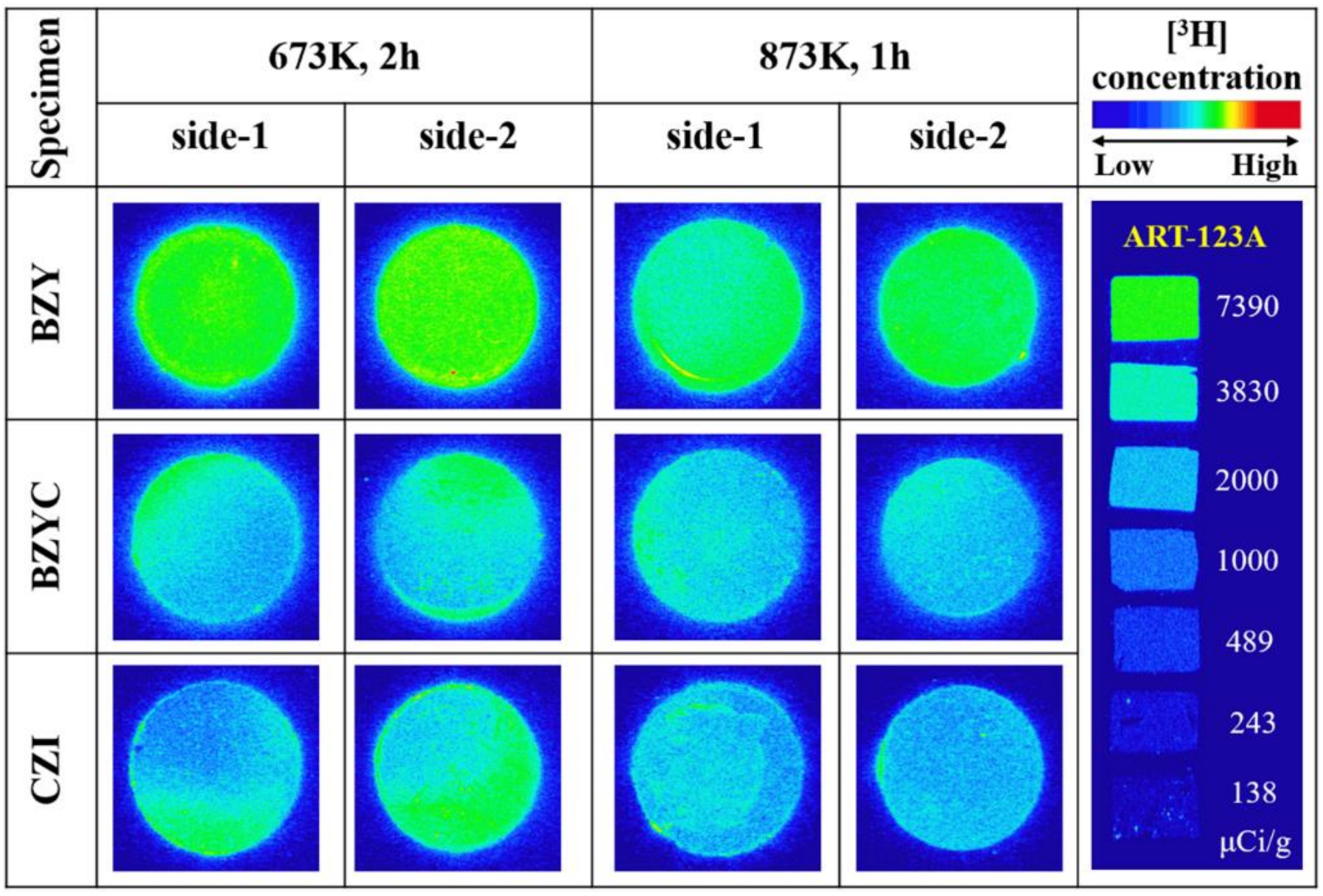

**Fig 6.** (Left) IP images of the outer surfaces (sides 1 and 2) of the BZY, BZYC, and CZI samples for different DTO-exposure conditions (673 K, 2 h and 873 K, 1 h). (Right) IP images for the tritium standard of ART-123A along with the corresponding radioactivity values. The color bar in the inset shows the tritium concentration. The IP exposure was carried out for 1 h at room temperature (298 K).

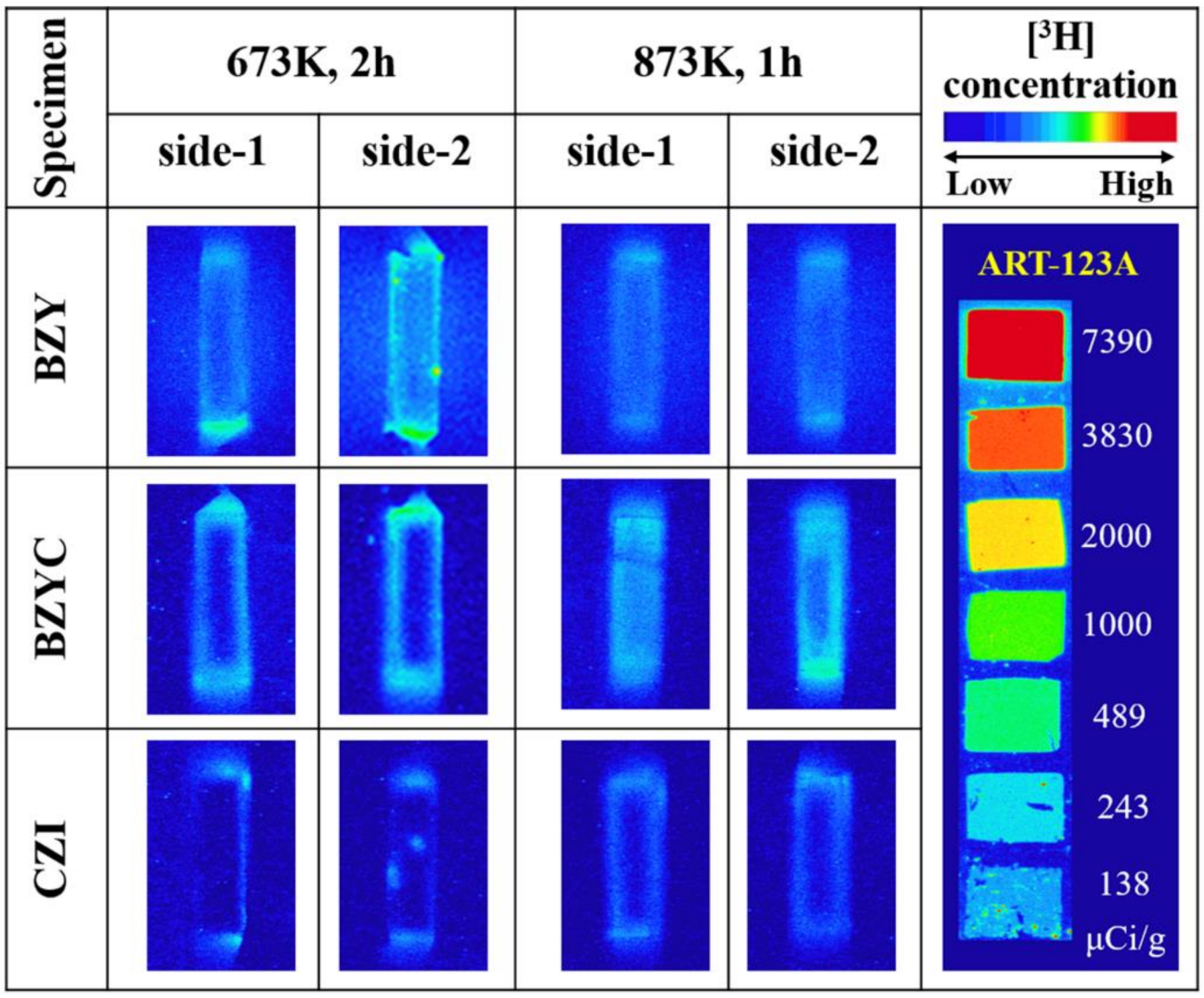

**Fig 7.** (Left) IP-images of the cross-sections (sides 1 and 2) of the cut BZY, BZYC, and CZI samples for different DTO-exposure conditions (673 K, 2 h and 873 K, 1 h). (Right) IP images for the tritium standard of ART-123A. The IP exposure was carried out for 17 h at room temperature (298 K).

## 4.2 Calculation of the hydrogen diffusivities from the IP image

The hydrogen diffusivities of the tritium-exposed samples were determined from the tritium concentration distribution profile obtained from the IP images. To explain the process of calculating the diffusivity from the cross-sectional tritium concentration distribution, we considered the CZI sample that was exposed to DTO at 873 K for 1 h. **Fig. 8(a)** shows the cross-sectional PSL intensity distribution for the CZI sample, and **Fig. 8(b)** shows the corresponding tritium concentration distribution profile obtained from the conversion of the PSL values. The PSL intensity distribution graph was based on the IP image of the sample's selected cross-section (**Fig. 8(a)**), and the data points were obtained by integrating the PSL values at the same depth from the surface to the center. The tritium concentration was highest at the surface of the sample, decreased toward the center, and was the lowest at the center. It was assumed that the top of the concentration distribution profile ($x$ = 1.225, 3.225) represented the sample surface; thus, a sample thickness of 2 mm was derived from the concentration distribution profile. This was lower than the actual sample thickness of 2.44 mm because tritium was released from the surface of the sample over time after the DTO-exposure.

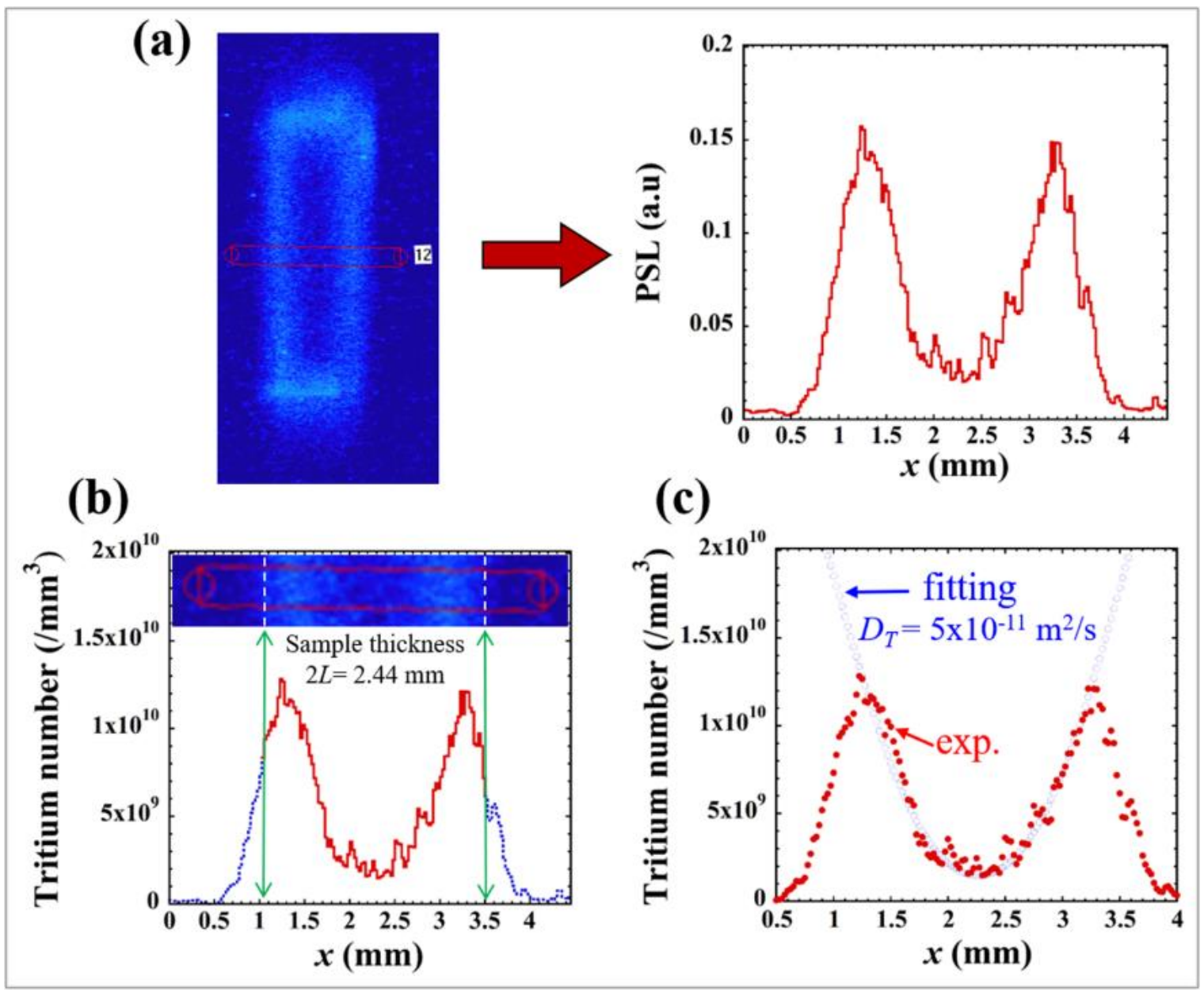

**Fig 8.** Calculation of the hydrogen diffusivity from the IP image of the DTO-exposed (873 K, 1 h) CZI sample: (a) IP image and the PSL intensity (arb. units) distribution curve, (b) tritium concentration distribution profile, and (c) tritium concentration distribution with the fitting curve.

Based on the solution of the diffusion equation, the fitting was performed by changing diffusivity as a fitting parameter for the tritium concentration distribution obtained from the IP image. Subsequently, the diffusivity of tritium in the sample was determined. It was assumed that tritium was equally released from the front and back of the sample during IP exposure. The midpoint of the two vertices in the tritium concentration distribution profile $C(x,t)$ was considered to be the middle of the sample thickness. Here, $x$ (in mm) is the tritium concentration at the sample surface and $t$ (in s) is the exposure time. Fitting was performed using a series solution of the diffusion equation by assuming the initial conditions (I.C.) and boundary conditions (B.C.) for the concentration profile $C(x,t)$ [52].

- Assumptions:

$$\mathrm{I.C.:}\ C(0,0) = C(2L,0) = 0$$

$$\mathrm{B.C.}\ : C(0,t) = C(2L,t) = C_s$$

- Series solution of the diffusion equation:

$$C(x,t) = C_s\left[\sum_{n=0}^{\infty}(-1)^n erfc\left\{\frac{(2n+1)L-x}{2\sqrt{D_T t}}\right\}\right] + C_s\left[\sum_{n=0}^{\infty}(-1)^n erfc\left\{\frac{(2n+1)L+x}{2\sqrt{D_T t}}\right\}\right] \quad (5)$$

In Eq. (5), $C(x,t)$ is the tritium concentration at a depth of $x$ (mm) from the sample surface, $2L$ is the width of the sample, $t$ (s) is the exposure time, and $D_T$ is the diffusivity of tritium. The boundary condition $C(x,t)$ was zero prior to exposure because tritium did not exist in the sample. The concentration at the sample surface ($C_s$) was fixed after starting the exposure. The fitting was performed by changing the values of $D_T$ and $C_s$. The value of $D_T$ was considered to best fit the actual tritium concentration distribution. When the actual fitting was performed according to Eq. (5), the left end of a $2L$-wide sample was located at $x = 0$. To match the tritium concentration distribution in the actual sample, Eq. (5) can be expressed as Eq. (6).

$$C(x,t) = C_s\left[\sum_{n=0}^{\infty}(-1)^n erfc\left\{\frac{(2n+1)L-(x-a)}{2\sqrt{D_T t}}\right\}\right] + C_s\left[\sum_{n=0}^{\infty}(-1)^n erfc\left\{\frac{(2n+1)L+(x-a)}{2\sqrt{D_T t}}\right\}\right] \quad (6)$$

$here, a =$ $x$ coordinate of the midpoint of the actual sample thickness

**Fig. 8(c)** shows the result of fitting the cross-sectional IP profile for the CZI sample (873 K, 1 h). The diffusivity of tritium in the CZI sample at 873 K was determined to be $5.0 \times 10^{-11}$ $m^2/s$. The diffusivities of tritium in the BZY, BZYC, and CZI samples at 673–873 K were calculated in the same manner by using the PSL profiles shown in **Figs. 9** and **10**.

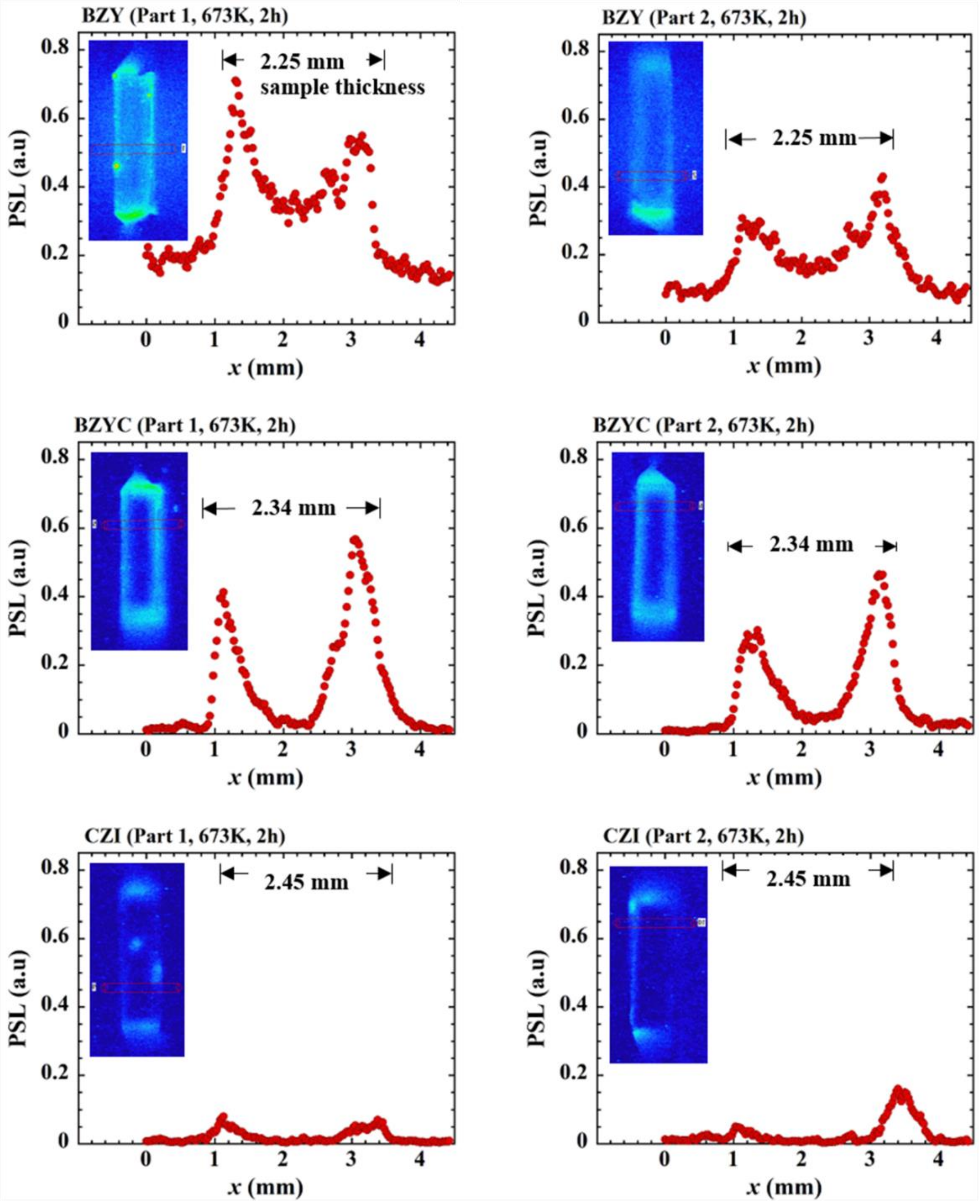

**Fig 9.** Line profiles of the IP images to calculate the hydrogen diffusivities for the DTO-exposed (673 K, 2 h) BZY, BZYC, and CZI samples. The y-axis represents the PSL intensity (arb. units), and the insets of each profile represent the corresponding IP images.

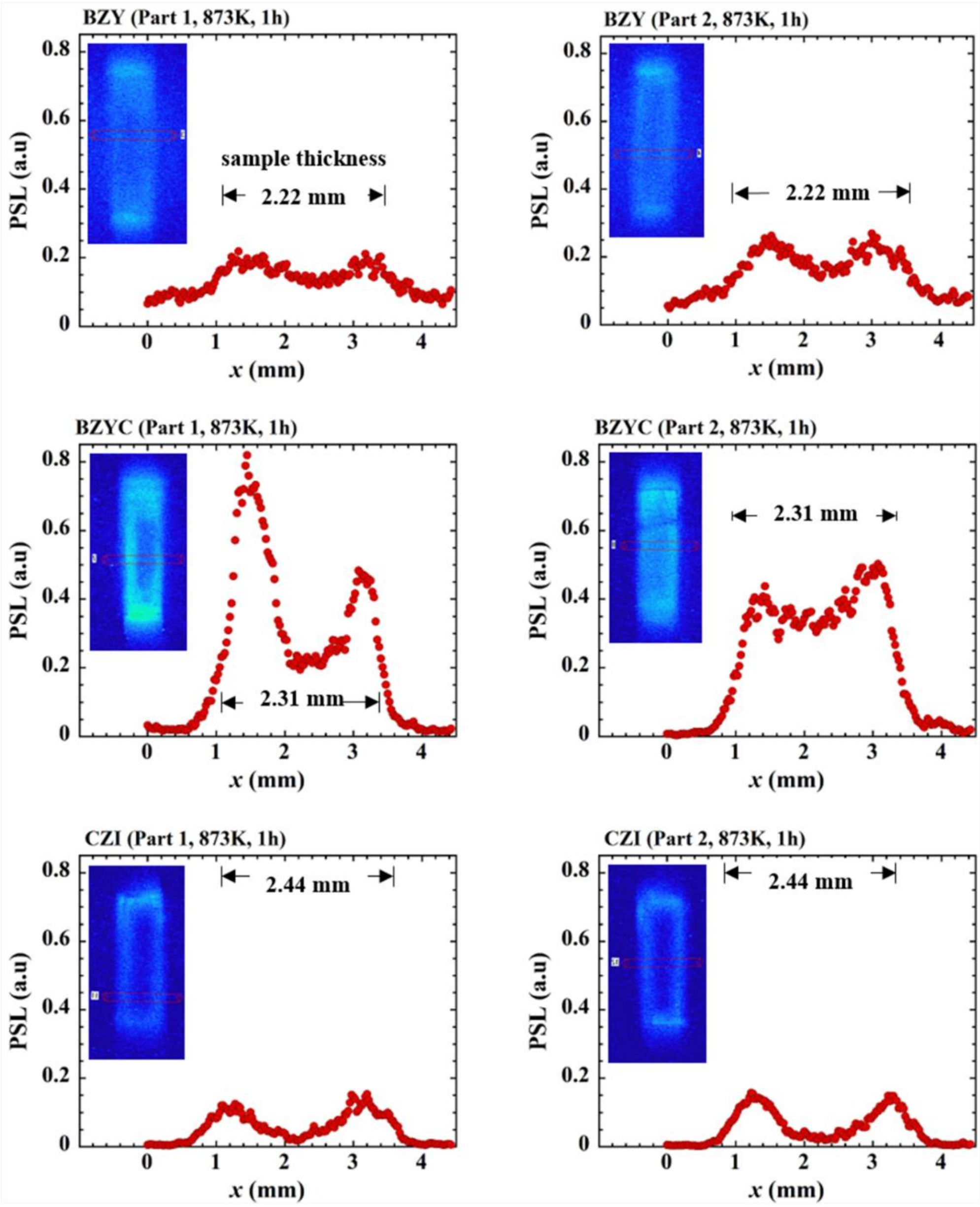

**Fig 10.** Line profiles of the IP images to calculate the hydrogen diffusivities of the DTO-exposed (873 K, 1 h) BZY, BZYC, and CZI samples. The y-axis represents the PSL intensity (arb. units), and the insets of each profile represent the corresponding IP images.

# 5 Results and Discussion

## 5.1 Visualization of tritium distribution

**Table 3** shows the tritium intensities (in PSL/mm$^2$·sec) at the outer surface and the near-surface cross-sections of the samples that were calculated from the IP images in **Figs. 6** and **7,** respectively. The tritium activities on the outer surfaces of the BZYC and CZI samples were similar; however, the tritium intensity (PSL/mm$^2$·sec) at the outer surface of the BZY sample was higher than that for the BZYC and CZI samples (**Table 3**). The near-surface cross-sectional tritium intensities for CZI were consistently lower than that for BZY and BZYC. Furthermore, the tritium intensities (PSL/mm$^2$·sec) at the outer surfaces were higher than that at the inner surfaces for all the samples. This was clearly demonstrated by the IP images (**Figs. 6** and **7**). These results indicated that the number of hydrogen atoms that were adsorbed on the sample surface was higher than the number of hydrogen atoms that were dissolved inside the samples. This phenomenon was attributed to the water dissolution or DTO vapor dissolution reactions on the surfaces of the proton conducting oxide materials [10,53–56].

**Table 3.** Summary of the experimental data obtained from the FLA7000 IP reader under the different DTO-exposure conditions.

| DTO-exposure temp. (K) | DTO-exposure time (h) | Specimen | Outer surface tritium intensity, PSL/mm$^2$.sec (arb. units) | Near-surface cross-sectional tritium intensity, PSL/mm$^2$.sec (arb. units) |
|---|---|---|---|---|
| 673 | 2 | BZY | (210 ± 20) x $10^{-4}$ | (4.8 ± 1.9) x $10^{-4}$ |
| | | BZYC | (84 ± 8) x $10^{-4}$ | (5.4 ± 0.8) x $10^{-4}$ |
| | | CZI | (90 ± 20) x $10^{-4}$ | (0.8 ± 0.4) x $10^{-4}$ |
| 873 | 1 | BZY | (140 ± 10) x $10^{-4}$ | (2.7 ± 0.2) x $10^{-4}$ |
| | | BZYC | (70 ± 5) x $10^{-4}$ | (6.4 ± 1.0) x $10^{-4}$ |
| | | CZI | (59 ± 5) x $10^{-4}$ | (1.4 ± 0.2) x $10^{-4}$ |

The cross-sectional IP images (**Fig. 7)** showed that tritium diffused deeper into BZY and BZYC than into CZI at 673–873 K. The non-uniform distribution of hydrogen was observed inside the BZY and CZI samples at a low exposure temperature (673 K). This was consistent with the results of our previous studies on BZY under HT-exposure conditions [13].

The non-uniform distribution of tritium was observed at the outer surfaces of both the BZYC and CZI samples at a low exposure temperature (673 K) (**Fig. 6**). This was inconsistent with the results of previous studies where a uniform distribution of tritium was observed at the outer surfaces of the BZYC samples [13]. Tritium was uniformly dissolved at the outer surfaces of the BZY samples under all exposure conditions (**Fig. 6).** However, the results of previous studies showed the non-uniform dissolution of tritium at the outer surfaces of the BZY samples at a low exposure temperature (673 K) [13]. The reason for this discrepancy might be that the DTO-exposure facilitated the uniform dissolution of tritium at the surface of and inside the BZY samples.

## 5.2 Hydrogen solubility

The amounts of dissolved hydrogen or hydrogen solubilities at 673 K and 873 K were calculated using the IP images of the cut surfaces (**Fig. 7**). **Table 4** shows the hydrogen solubilities and diffusivities in all the samples along with the corresponding error at each exposure temperature. **Fig. 11** shows the Arrhenius plot of the hydrogen solubilities for all the samples. The data in the figure were obtained by integrating the amount of hydrogen near the surfaces of the samples where the dissolved amount of hydrogen had reached a steady level. The error bar indicates the standard deviation. The amount of dissolved hydrogen in the BZY samples decreased with the increase in the exposure temperature; however, the amount of dissolved hydrogen in the BZYC and CZI samples increased with the increase in the exposure temperature (**Fig. 11).** Since the degree of error was low at high exposure temperatures, there was a uniform distribution of the hydrogen concentration at the same depth from the sample surface at 873 K. Since the error was high at low exposure temperatures, there was non-uniform distribution of the hydrogen concentration at the same depth from the sample surface that was confirmed by the IP images (**Fig. 11)**.

**Table 4.** Summary of the hydrogen solubilities and diffusivities for BZY, BZYC, and CZI under different DTO-exposure conditions.

| Exposure temperature (K) | Exposure time (h) | Specimen | Thickness (mm) | H-solubility H mol/ 1 mol of oxides | H-diffusivity (m$^2$/s) |
|---|---|---|---|---|---|
| 673 | 2 | **BZY** | 2.25 | (1.0 ± 0.6) x $10^{-3}$ | (3.78 ± 0.02) x $10^{-11}$ |
| | | **BZYC** | 2.34 | (8.2 ± 2.0) x $10^{-4}$ | (1.6 ± 0.2) x $10^{-11}$ |
| | | **CZI** | 2.45 | (9.7 ± 4.0) x $10^{-5}$ | (6.5 ± 2.5) x $10^{-12}$ |
| 873 | 1 | **BZY** | 2.22 | (5.1 ± 0.08) x $10^{-4}$ | (9.4 ± 0.1) x $10^{-11}$ |
| | | **BZYC** | 2.31 | (1.1 ± 0.2) x $10^{-3}$ | (6.25 ± 0.75) x $10^{-11}$ |
| | | **CZI** | 2.44 | (2.6 ± 0.2) x $10^{-4}$ | (5.5 ± 0.5) x $10^{-11}$ |

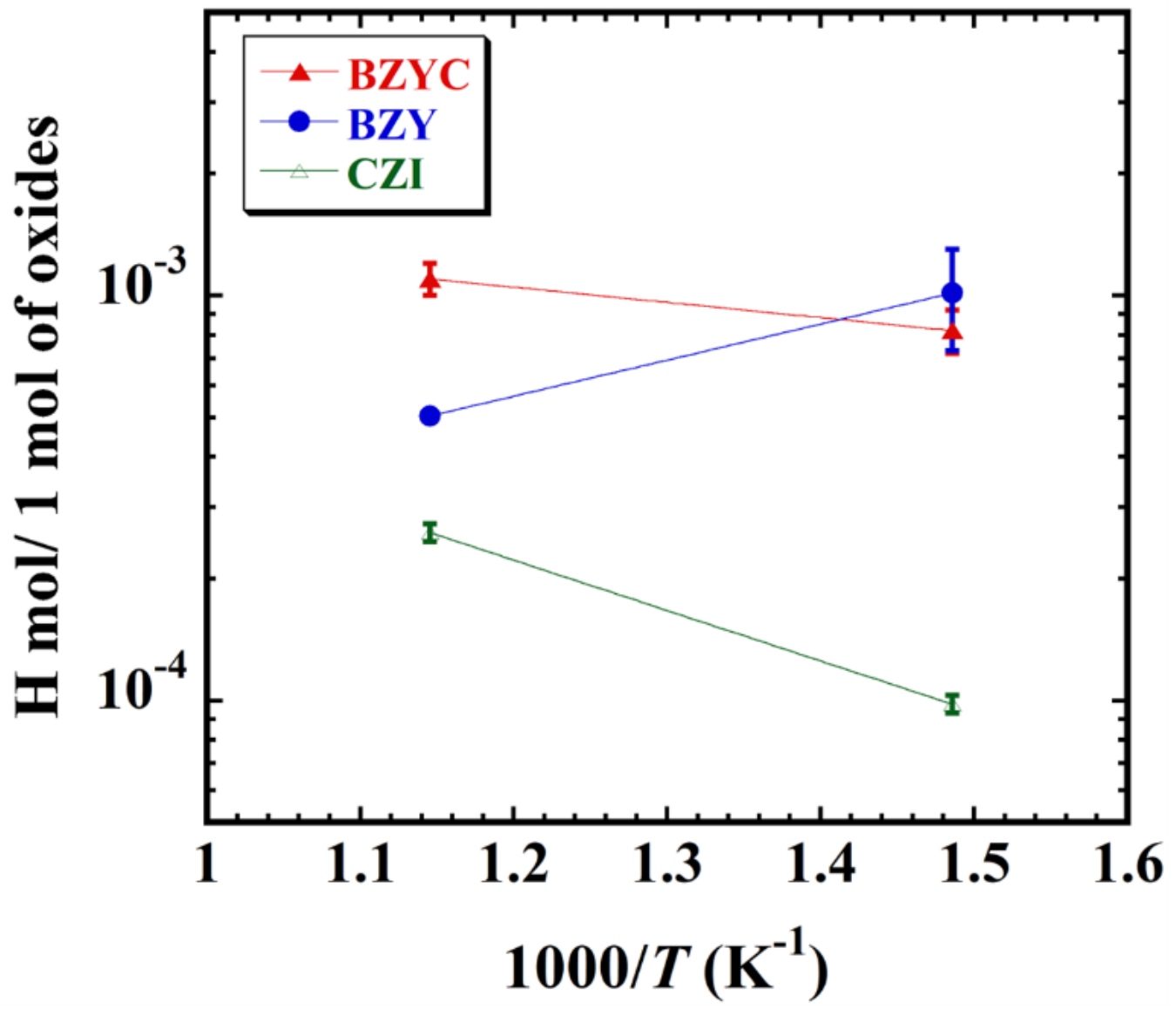

**Fig 11.** Arrhenius plot of the hydrogen solubilities for the BZY, BZYC, and CZI samples (2.3 kPa; 673–873 K). The y-axis represents the molar fraction of hydrogen and the oxide samples.

In this study, the hydrogen solubility in CZI was consistently lower than that in BZY and BZYC. The hydrogen solubility in proton-conducting oxides is directly proportional to the protonic conductivity [10,57] i.e., if a proton conductor shows high hydrogen solubility, then it also shows high proton conductivity and vice-versa. The experimental data that showed the temperature dependence of the hydrogen solubility was consistent with the literature data of the proton conductivity in barium and calcium zirconates that was reported by H. Iwahara *et al.* [58]. But an opposite trend in the temperature dependence of the hydrogen solubility was observed in our previous studies for the HT-exposed BZY and BZYC samples [13]. The hydrogen solubility increased with the increase in the exposure temperature for the HT-exposed BZY sample while it decreased with the increase in the exposure temperature for the HT-exposed BZYC sample [13]. In current experiment tritium water (DTO) vapor was used as the tritium source, where tritiated water plays major role for hydrogen or tritium dissolution through the samples. On the other hand, in our previous paper [13] tritium gas (HT gas) was used as the tritium exposure source. As the exposure sources are different, we may assume that tritium dissolution through the samples are different. Oxygen vacancies are generated in the BZY, BZYC, and CZI perovskite oxides by the substitution of the B-site Zr cations with Y, Co, or In. These vacancies provide a path for the dissolution of hydrogen or hydrogen isotopes like deuterium and tritium. Since the samples were exposed to DTO vapor, there was a possibility of a reaction between $D_2O$ and the oxygen vacancies. The reactions for the creation of oxygen vacancies in BZY and water incorporation are presented by the following Kröger-Vink notation equations (Eqs. 7 and 8) [10,53–56] that are shown below.

$$\mathrm{Y_2O_3} + 2\mathrm{Zr}_{\mathrm{Zr}}^{\mathrm{x}} + \mathrm{O}_{\mathrm{o}}^{\mathrm{x}} \leftrightarrow 2\mathrm{Y}_{\mathrm{Zr}}' + \mathrm{V}_{\mathrm{o}}^{\bullet\bullet} + 2\mathrm{ZrO_2} \quad (7)$$

$$\mathrm{D_2O} + \mathrm{V}_{\mathrm{o}}^{\bullet\bullet} + \mathrm{O}_{\mathrm{o}}^{\mathrm{x}} \leftrightarrow 2\mathrm{OD}_{\mathrm{o}}^{\bullet} \quad (8)$$

Eq. (8) indicates that $\mathrm{V}_{\mathrm{o}}^{\bullet\bullet}$ is an oxygen ion vacancy with two positive charges. This vacancy reacts with one $\mathrm{D_2O}$ molecule and one oxygen anion to produce two OD ions. Therefore, the replacement of the oxygen vacancies with protonic defects in the form of OD ions facilitates the incorporation of water in the BZY perovskite proton-conducting materials in case of DTO exposure. On the other hand, in the previous studies we used the HT gas as the tritium source, there are possibility to occurs the reaction between the HT gas with $O_o^x$ ion and produce an oxygen ion vacancy ($\mathrm{V}_{\mathrm{o}}^{\bullet\bullet}$) which help to dissolution of HT gas through the proton conducting materials. This possible reaction of HT gas or $H_2$ gas with $O_o^x$ ion is presented by the following Kröger-Vink notation equation (Eq. 9) where $O_o^x$ means one oxygen ion in the oxygen site with neutral charge.

$$H_2\,(g) + O_o^x \leftrightarrow H_2O\,(g) + V_o^{\bullet\bullet} + 2e' \quad (9)$$

Moreover, in current experiment at lower exposure temperature (673 K) both BZY and BZYC showed almost similar order of hydrogen solubilities i.e. ~$10^{-3}$ (H/M), but in the previous paper [13] this order was about one order higher in the case of BZYC than the BZY. In the previous paper [13] at high temperature (873 K), BZYC showed higher hydrogen solubility than the BZY which is consistent with current paper's solubilities data at higher temperature. Therefore, we may assume that in both experiments, BZYC shows higher hydrogen solubilities which may be due to the participation of cobalt (Co) content in the $BaZrO_3$ perovskite as a catalyst to enhance the hydrogen dissolution during tritium exposure [38,39,59].

## 5.3 Hydrogen diffusivity

**Fig. 12** shows the comparison of the Arrhenius plots of the hydrogen diffusivities in all the samples with the literature data of the protonic diffusivity in BZY. A clear temperature dependence was demonstrated by the plot because the diffusivity increased with the increase in the exposure temperature. The approximate line formed by the data points with the error bar was used to frame the equations to determine the diffusivity of each sample.

$$D_{\mathrm{T\,(BZY)}}\ [\mathrm{m^2/s}] = 2.0 \times 10^{-9} \exp\left(-\frac{0.23\ \mathrm{eV}}{k_B T}\right) \quad (10)$$

$$D_{\mathrm{T\,(BZYC)}}\ [\mathrm{m^2/s}] = 6.1 \times 10^{-9} \exp\left(-\frac{0.31\ \mathrm{eV}}{k_B T}\right) \quad (11)$$

$$D_{\mathrm{T\,(CZI)}}\ [\mathrm{m^2/s}] = 7.3 \times 10^{-8} \exp\left(-\frac{0.54\ \mathrm{eV}}{k_B T}\right) \quad (12)$$

where $k_B$ is the Boltzmann constant and $T$ is the exposure temperature of the sample. The addition of Co affected the hydrogen diffusivity in the BZY samples. The activation energy ($E_a$) of diffusion for the BZY, BZYC, and CZI samples was 0.23 eV, 0.31 eV, and 0.54 eV, respectively.

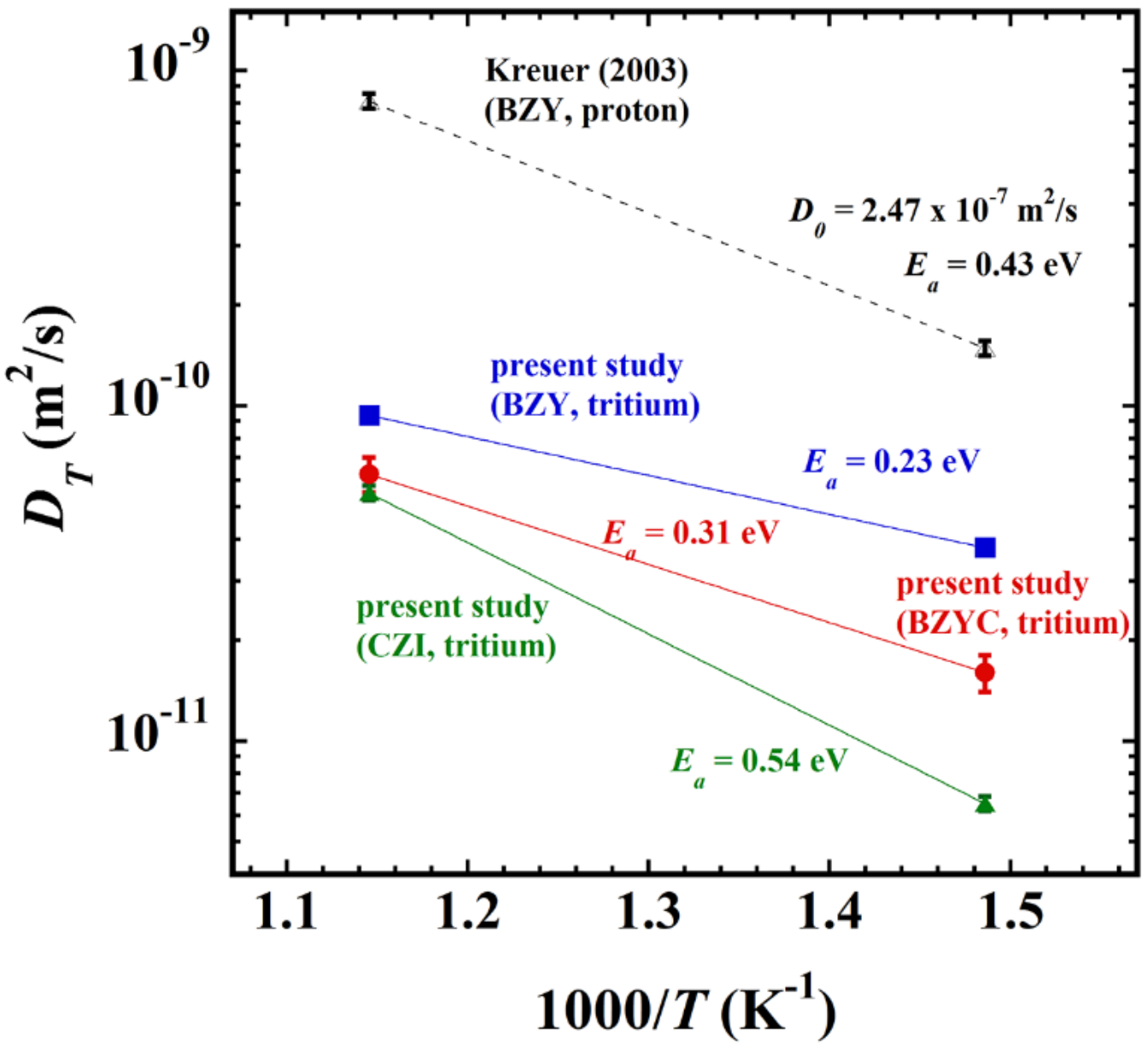

**Fig 12.** Arrhenius plot of the hydrogen diffusivities to compare the tritium diffusivity data for BZYC (closed red circle), BZY (blue square), and CZI (green triangle) that were obtained from the present study with the literature data of the proton diffusivity in BZY at 673–873 K that was reported by Kreuer [15] (open triangle).

$E_a$ of diffusion is significantly affected by the interaction of oxygen and hydrogen with the B-site cations (Zr, Y, In, and Co) of the perovskite structure [60]. The interaction of oxygen and hydrogen with Co is weaker than that with Zr and Y [38]. Therefore, $E_a$ of diffusion for BZY (0.74 eV) was higher than that for BZYC (0.18 eV) due to the stronger interactions in BZY as compared to that in BZYC, reported in our previous study for HT-exposure case [13]. But current experiments show different results, i.e. activation energies of diffusion for BZY (0.23 eV) is lower than the BZYC (0.31 eV). One reason for this difference might be that as the exposure tritium are different for current experiment, the tritium dissolution mechanism through the samples might be different which might be affected the activation energies of diffusion.

Moreover, the low $E_a$ (0.23 eV) of diffusion in BZY that was obtained in the present study was consistent with the literature data [15,61], the pre-exponential factor ($D_0 = 2.0 \times 10^{-9}$) itself was low [15]. These variations could be attributed to multiple factors, one of which is that the nature of the diffusing atoms in this experiment and the literature was different, i.e., tritium, and proton, respectively. This indicated that the diffusion speed was different due to the isotopic effect [56]. The diffusivity $D$ can be expressed as $D = 1/\sqrt{m}$, where m is the mass of the diffused atom. Since the mass of tritium is three times the mass of a hydrogen proton, the diffusivity of tritium is $1/\sqrt{3}$ times that of a proton [10,11]. The hydrogen diffusivity in CZI was consistently low. This was consistent with the solubility data obtained from the present experiment.

# 6 Conclusions

In this study, the hydrogen solubility and diffusivity of proton-conducting calcium and barium zirconates were measured using a tritium imaging plate. The hydrogen distribution at the surface and inside of the samples was determined from the IP images. The amount of dissolved hydrogen was calculated using the tritium standard sample by converting the PSL value to the tritium number. The temperature dependence of the amount of dissolved hydrogen was different for the BZY, BZYC, and CZI samples. The diffusivity of tritium in the calcium and barium zirconates was determined from the cross-sectional tritium distribution profile that was obtained from the IP images. The hydrogen diffusivity was affected by the presence of Co that was added as a sintering aid; moreover, the activation energy of hydrogen diffusivity in CZI was high. The results for all the samples were compared. It was concluded that under DTO-exposure at 673–873 K, the hydrogen solubilities and diffusivities in the barium zirconates were higher than that in the calcium zirconates.

## Data availability

The raw/processed data required to reproduce these findings cannot be shared at this time as the data also forms part of an ongoing study.

## CRediT authorship contribution statement

**M. Khalid Hossain:** Conceptualization, Data curation, Formal analysis, Investigation, Methodology, Software, Validation, Visualization, Project administration, Writing - original draft, Writing - review & editing. **K. Hashizume:** Conceptualization, Funding acquisition, Project administration, Resources, Supervision, Validation, Visualization, Writing - review & editing. **Y. Hatano:** Project administration, Resources.

## Declaration of competing interest

The authors declare that they have no known competing financial interests or personal relationships that could have appeared to influence the work reported in this paper.

## Acknowledgements

Authors are grateful to the Hydrogen Isotope Research Center (HIRC), University of Toyama for the support to carry out tritium experiments.

# Authors biography

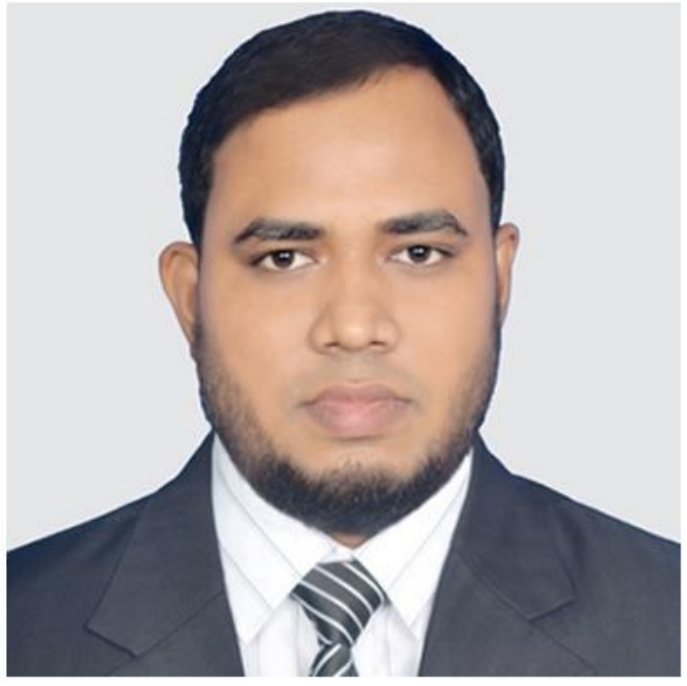

**M. Khalid Hossain** is presently engaged in research on nuclear energy materials as an MEXT fellow at Kyushu University, Japan. He has worked as a Research Scientist at the Bangladesh Atomic Energy Commission since 2012. M. Khalid Hossain is a Life Member of the Bangladesh Physical Society, Bangladesh Electronic Society, and Bangladesh Solar Energy Society among others. He is also a member of the Bangladesh Atomic Energy Scientist Association. His research interests include energy materials, functional materials, perovskite proton-conducting oxides, thin films, and photovoltaic devices among others. He has authored and co-authored 33 SCI(E) articles that have been published in reputable peer-reviewed journals.

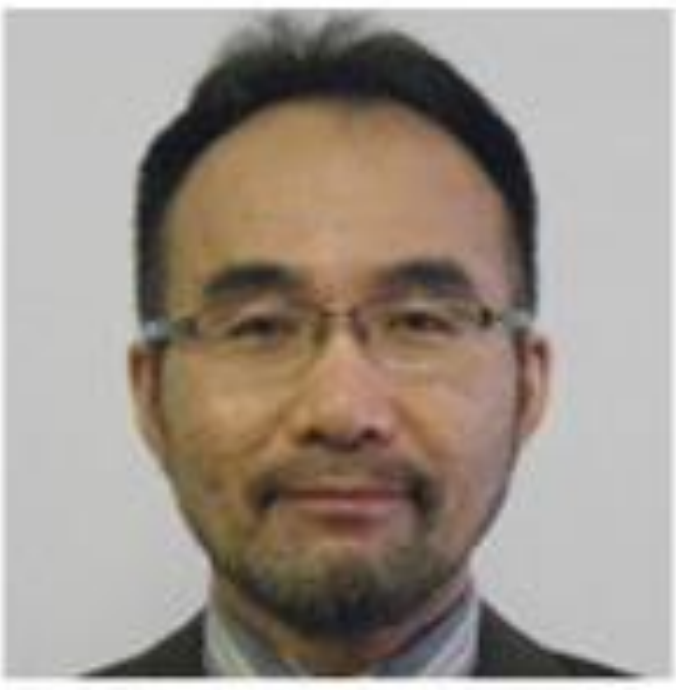

**Kenichi Hashizume** is an Associate Professor in the Dept. of Advanced Energy Engineering Science, Kyushu University, Japan. He graduated from the Department of Nuclear Engineering at Kyushu University. His present research interests include nuclear materials, energy-related materials, and the interactions between hydrogen and materials. His potential research themes include the hydrogen behavior in alloys and ceramics, control materials for fission reactors, and applications of radiation energy among others. He is a member of the Japan Society of Plasma Science and Nuclear Fusion Research, Atomic Energy Society of Japan, Japan Institute of Metals, Solid State Ionics Society of Japan, and Ceramic Society of Japan.

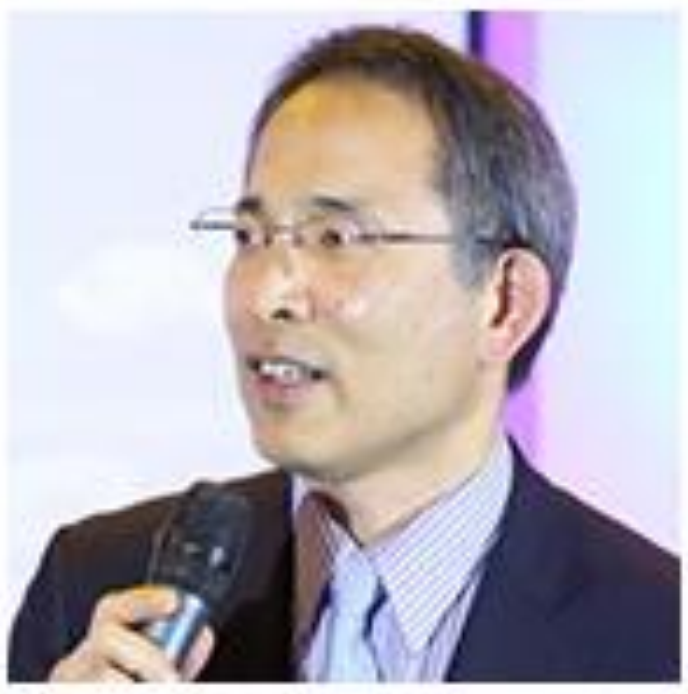

**Yuji Hatano** is a Professor at the Hydrogen Isotope Research Center, Organization for Promotion of Research, University of Toyama, Japan. He graduated from the Department of Nuclear Engineering at Kyushu University in 1989. His research interests include fusion engineering, hydrogen energy, materials science, and tritium science. He has authored and co-authored more than 200 SCI(E) articles that have been published in reputable peer-reviewed journals.